\documentclass{aa}  

\usepackage{graphicx}
\usepackage{txfonts}
\usepackage{acronym}
\usepackage{xcolor} 
\usepackage{booktabs}
\usepackage{longtable}

\usepackage{hyperref}
\hypersetup{
  colorlinks=true,        
  linkcolor=blue,         
  citecolor=cyan,         
}
\usepackage{natbib}

\newcommand{\sgra}{Sgr A$^*$}
\newcommand{\m}[1]{M\,87$^*$}

\newacro{SMBH}{supermassive black hole}
\newacro{BH}{black hole}
\newacro{DM}{dark matter}
\newacro{GRHD}{general relativistic hydrodynamics}
\newacro{GRMHD}{general relativistic magnetohydrodynamics}
\newacro{BHAC}{black hole accretion code}
\newacro{EHT}{Event Horizon Telescope}
\newacro{NFW}{Navarro-Frenk-White}
\newacro{AGN}{active galactic nuclei}
\newacro{QSO}{quasi-stellar object}
\begin{document} 

\title{General relativistic quasi-spherical accretion in a dark matter halo}
\titlerunning{Accretion in a DM halo}
\author{Razieh Ranjbar \thanks{ORCID: https://orcid.org/0000-0003-1488-4890}\inst{1}
        \and Héctor R. Olivares-Sánchez \thanks{ORCID: https://orcid.org/0000-0001-6833-7580} \inst{2}}

\institute{ School of Astronomy, Institute for Research in Fundamental Sciences, Tehran, Iran \\ \email{r.ranjbar@ipm.ir}
\and Departamento de Matem\'{a}tica da Universidade de Aveiro and Centre for Research and Development in Mathematics and Applications (CIDMA), Campus de Santiago, 3810-193 Aveiro, Portugal \\
\email{h.sanchez@ua.pt}}

   \date{\today}

  \abstract
   {The Bondi spherical accretion solution has been used to model accretion onto compact objects in a variety of situations, from interpretation of observations to subgrid models in cosmological simulations.}
   {We aim to investigate how the presence of \ac{DM} alters the dynamics and physical properties of accretion onto supermassive black holes on scales ranging from $\sim 10$ pc to the event horizon.}
   {In particular, we investigate Bondi-like accretion flows with zero and low specific angular momentum around supermassive black holes surrounded by dark-matter halos by performing 1D and 2.5D \ac{GRHD} simulations using the \ac{BHAC}.}
   {We find notable differences in the dynamics and structure of spherical accretion flows in the presence of \ac{DM}. The most significant effects include increases in density, temperature, and pressure, as well as variations in radial velocity both inside and outside the regions containing \ac{DM} or even the production of outflow.}
  {This investigation provides valuable insights into the role of cosmological effects, particularly \ac{DM}, in shaping the behavior of accretion flows and \acp{BH}. Our simulations may be directly applicable to model systems with a large black hole-to-halo mass ratio, which are expected to be found at very high redshifts.}


   \maketitle
%

\section{Introduction}

The heart of each galaxy is home to a \ac{SMBH}. These black holes grow when nearby gas and dust are swallowed by their deep gravitational wells. Despite the evident potential of \acp{SMBH} in characterizing objects at galactic centers, many aspects of their physics remain unknown. For example, existing models attempting to explain the origins of \acp{SMBH} fail to accurately predict their observed distribution in mass and distance \citep{Kormendy2013SMBHGalaxies}. One of the key questions with cosmological implications is how they increased in mass so quickly. Observed spectra can provide considerable information on the accretion systems \citet{Abramowicz1988}. However, solely focusing on luminous matter components such as gas, dust, and the accretion disk may not be sufficient to explain this observed high growth rate.

It is believed that $85 \%$ of the Universe's mass is invisible \ac{DM} \citep{Jarosik2011WMAP}. Based on numerous observations and N-body simulations, it is now widely accepted that galactic centers contain a \ac{SMBH}, enclosed within a \ac{DM} halo that extends at much larger distances for millions of light-years \citep[see,e.g.][]{deMartino2020DarkMatter}. The presence of \ac{DM} in galaxies, supported by substantial evidence such as the rotation curve of galaxy disks \citep{Battaglia2006}, suggests that the \ac{DM} halos surrounding galaxies likely influence black holes and the accreting gas. \ac{DM} content in elliptical and disk galaxies can be measured with the weak gravitational lensing of background galaxies. The most massive black holes are found at the center of galaxies and grow during accretion phases, observed as \ac{AGN}. Despite advances in understanding \ac{AGN} properties and host galaxies, their connection to large-scale cosmic environments and \ac{DM} halos remains unclear. Understanding how \ac{AGN} activity is affected by these environments can provide important insights into SMBH fueling and feedback mechanisms (\citet{Powell2022BASS}). In this study, we aim to contribute to the understanding of the effects of the \ac{DM} environment on the growth and coevolution of a galaxy's central \ac{SMBH} by studying the problem of spherical and quasi-spherical accretion onto a black hole surrounded by a \ac{DM} halo. In particular, we focus on the case of accretion onto \acp{SMBH} at the centers of elliptical galaxies.

When modeling the accretion flow in the close vicinity of compact objects such as \acp{SMBH},
it becomes crucial to consider the gravitational field in a general relativistic fashion,
with the spacetime around the object being described by a metric tensor, which is a solution of the Einstein field equations.
The simplest spacetime geometry that captures this scenario is that of a static, spherically symmetric \ac{BH}, i.e., the
Schwarzschild solution. However, when describing the spacetime around a black hole surrounded by dark matter, it becomes necessary to consider non-vacuum solutions. The dynamics of the accretion flow can be significantly altered when different distributions of matter are present outside the black hole. Due to this, different \ac{DM} models may have different observational signatures.
Through different observations and simulations, we have a rough understanding of the distribution of \ac{DM} in galaxies.
Some of the most popular \ac{DM} profiles in the literature are those of Navarro-Frenk-White \citep{Navarro1996CDMHalos}, Hernquist \citep{Hernquist1990GalaxyModel}, Jaffe \citep{Jaffe1983GalaxyModel}, and King \citep{King1962StarClusters}.

Bondi accretion \citep{1952MNRAS.112..195B}
and its relativistic generalization by
Michel \citep{Michel1972} represent
the conceptually simplest case of accretion -- that of a spherically symmetric object in a uniform medium.
However, both the original model and
some of its extensions are not only intrinsically interesting from a mathematical physics point of view,
but also possess important astrophysical applications.
For example, the growth of Primordial Black Holes can be inferred as a sequence of nearly stationary accretion stages of spherical accretion of ultrarelativistic gas \citep{LoraClavijo2013PBHAccretion}. Also, spherical accretion is important at the early stage of expansion in cosmological models \citep{Zeldovich1967}.
A simple extension of the model
for a homogeneous medium expanding uniformly
has been used to model accretion onto a point mass at the center of supernova explosions \citep{Colpi1996}.
The formation, accretion, and growth of \acp{SMBH} in the early universe have been investigated using the Bondi accretion rate onto initial seed \acp{BH} with masses $ M_{BH} \sim 10^2 - 10^3 M_{\odot}$, that can evolve into a \acp{SMBH} with masses $ M_{BH}  \sim 10^9 - 10^{10} M_{\odot}$ \citep{Moffat2020SMBHAccretion}. Also, Bondi accretion can be used to calculate and estimate the mass of intermediate-mass \acp{BH} \cite{Wrobel2018}. In cosmological simulations, Bondi prescriptions are often used to estimate the rate at which gas accretes onto \acp{SMBH} \citep{2019MNRAS.486.2827D, Ricotti2007BondiAccretion}. 

In this model,
accretion can be considered to start at the Bondi radius, where the gravitational potential of the \ac{SMBH} dominates the thermal energy of the surrounding hot gases \citep{Russell2015}. The \ac{EHT} observed the \acp{SMBH} at the core of the galaxy \m{87} (from now on \m{87}) and the Milky Way (\sgra),
producing images of the accreting plasma at the event-horizon scales \citep{EHT2019M87Shadow, EventHorizon2023SagittariusA}. The mass of \m{87} is  $6.5 \times 10^9 M_{\odot}$, corresponding to a Bondi radius of approximately $r_B \sim 0.12$  kpc. \sgra has a mass of $4.1 \times 10^6\ M_{\odot}$, and a Bondi radius of $ \sim 0.1$ pc \citep{Falcke2013EventHorizon}. Observations suggest that \m{87} possess a \ac{DM} halo with a mass of approximately $10^{14}\ M_{\odot}$ with a characteristic scale of about $100$ kpc \citep{DeLaurentis2022M87}. Similarly, \sgra has an estimated halo mass of $1.9 \times 10^{11}\ M_{\odot}$ within a radius of $20$ kpc \citep{Posti2019MWDarkMatterHalo}. In recent years, Bondi accretion has been generalized to include the effects of the gravitational field of the host galaxy with a central black hole and \ac{DM} halo in pseudo-Newtonian gravity, while still keeping the problem analytically tractable \citep{Ciotti2018JaffeModels, Mancino2022BondiAccretion}.

In this study, we investigate the scenario of a nonrotating \ac{BH} surrounded by \ac{DM} that is accreting from a uniform medium. \ac{DM} deforms the Schwarzschild geometry sufficiently to alter the structure of the accretion flow around the \ac{SMBH}. In particular, we concentrate on \ac{DM} distributed following the Hernquist and \ac{NFW} profiles.
We first perform a detailed comparison between the Newtonian (Bondi 1952) model and its general relativistic generalization by \citet{Michel1972} by studying the analytical solution. Then, using general relativistic hydrodynamic simulations, we study spherical accretion from large ($\sim10$ pc) scales onto the vicinity of the \ac{BH}'s event horizon, exploring the influence of the \ac{DM} halo on the mass accretion rate, and the structure of the flow. We find that the effect of \ac{DM} becomes more significant as the halo mass $M_{halo}$ and characteristic scale $a_0$ increase. Finally, we extend Michel’s solution by considering a small amount of gas rotation in addition to the influence of the \ac{DM} halo.

The paper is organized as follows. In section \ref{sec.2}, we recall the main properties of the classical Bondi and Michel models. In section \ref{sec.3}, we introduce the dark matter density profile and derive the spherically symmetric metric of a \ac{BH} surrounded by a \ac{DM} halo. In section \ref{Setup} we briefly discuss the numerical setup. In section \ref{result}, we discuss the results of 1D and 2.5D simulations. Finally, we summarize and discuss our results in section \ref{sec.5}.

\section{Semi-analytic stationary solutions}\label{sec.2}

\subsection{Bondi Model}
A spherical symmetric steady-state accretion flow (with zero angular momentum and no magnetic field) onto a gravitating body was first described by (Bondi 1952) under Newtonian physics and is often called the Bondi Model. This model is foundational in understanding simple accretion processes. The flow is supposed to be steady and one-dimensional in the radial ($r$-)direction. The flow is further assumed to be inviscid and adiabatic, and magnetic and radiation fields are ignored. These assumptions limit the applicability of the simplest model to environments where these effects are minimal.
Under the Newtonian approximation, the continuity equation and the equation of motion are, respectively, 

\begin{equation}
	\frac{1}{4\pi r^2} \frac{d}{dr}(4\pi r^2 \rho v)= 0, 
\end{equation}

\begin{equation}
	v\frac{dv}{dr} = - \frac{1}{\rho} \frac{dp}{dr} - \frac{d\Phi}{dr},
\end{equation}
where $v$ is the flow velocity (negative for accretion), $\rho$ is the density, $p$ is the pressure, and $\Phi(r)$ is the gravitational potential. The polytropic relation $p = K\rho^\gamma$ is assumed, where $K$ and $\gamma$ are constants.

While in the classical Bondi model, $\Phi$ represents the potential of a point source, the inclusion of dark matter introduces additional gravitational effects that alter the density and velocity profiles.
We can go beyond the original model and have $\Phi$ including the potential of a black hole and a dark matter halo.
\begin{equation}
	\Phi = \phi_{BH} + \phi_{DM}
\end{equation}
Integrating the above equations yields the continuity equation and the Bernoulli equation:

\begin{equation}
    \label{eq:mdot-newtonian}
	- 4 \pi r^2 \rho v = \dot{M},
\end{equation}

\begin{equation}
    \label{eq:bernoulli-newtonian}
	\frac{1}{2} v^2 + \frac{\gamma}{\gamma -1} \frac{p}{\rho} + \Phi =  E,
\end{equation}
where $\dot{M}$ is the mass accretion rate and $E$ is the Bernoulli constant. 
By considering the regularity condition at the sonic point, $ v_c = c_{s,c} $, and fixing the gravitational potential, we can obtain the location of the sonic point and the speed sound, and find the radial profiles for all of the hydrodynamic variables.

\begin{table}[t]
	\centering
	
	\begin{tabular}{ |p{2cm}|p{4cm}| }
		\hline
		$(\alpha, \beta, \gamma)$ & Name \\
		\hline
		(2, 5, 0) & Plummer Profile  \\
		(2, 4, 0) & Perfect Sphere  \\
		(2, 3, 0)  & Modified Hubble Profile  \\
		(2, 3, 0)  & Modified Isothermal Sphere \\
		(1, 3, 1) & \ac{NFW} Profile  \\
		(1.5, 3, 1.5) & Moore Profile \\
		(1, 4, 1)  & Hernquist Profile \\
		(1, 4, 2)   & Jaffe Profile \\
		\hline
		
	\end{tabular}
	\caption{List of double power-law density profiles defined by parameters \((\alpha, \beta, \gamma)\) and their corresponding names.}
	\label{table1}
\end{table}

\subsection{Mass accretion rate}

The mass-accretion rate plays a crucial role in accretion disk theory as it governs the accretion mode of black holes. In large-scale cosmological simulations, the Bondi accretion formula is commonly employed to estimate the black hole's accretion rate. It has also been found that dark matter is a major contributor to black hole accretion physics and that its influence can be seen in the variation of the black hole's central energetics as an increase in the mass accretion rate. The Bondi accretion rate, which characterizes the mass accretion rate of transonic spherical accretion onto compact objects like black holes, is determined by the gas density and temperature at the outer boundary in its Newtonian form. By defining the Bondi radius as $r_B = GM/c_{s,\infty}^2$ (the sound speed $c_{s,\infty}$ of the ambient gas determined by its temperature), we can express the Bondi accretion rate as $ \dot{M}_{B} = \Lambda \gamma^{-3/2} 4 \pi r_B^2 \rho_{\infty} c_{s,\infty}$. For $ \gamma= 5/3$, the constant $ \Lambda(\gamma) $ is $0.25$, while for $\gamma = 1$, it is $1.12$ \cite{Park2009AccretionRate}.
Recent \ac{EHT} results targeted the \m{87} black hole and the center of our galaxy, revealing images of \m{87} and \sgra, released in 2019 and March 2024, respectively. The accretion rate of \m{87} is estimated to be \( \dot{M} \sim (3-20) \times 10^{-4} \, M_{\odot} \, \text{yr}^{-1} \) \cite{EHT2019M87Shadow}, which is about a hundred times smaller than the Bondi accretion rate prediction. Also, the accretion rate of \sgra in simulations from \cite{2023MNRAS.521.4277R} falls within the range \( 2.5 \times 10^{-9} - 2 \times 10^{-8} \, M_{\odot} \, \text{yr}^{-1} \), consistent with previous estimates of the accretion rate onto \sgra and with estimates based on the EHT-observed flux.
Chandra observations using ACIS-I, centered on the position of Sagittarius A* (\sgra), indicate that its accretion rate may be 1–2 orders of magnitude lower than the estimated Bondi rate, which is approximately \( \sim 10^{-7} - 10^{-8} \, M_{\odot} \, \text{yr}^{-1} \) \cite{Baganoff2003}. While Bondi accretion is mathematically simple, it is insufficient for accurately describing real astrophysical systems, necessitating several modifications. One significant but less frequently discussed factor is the influence of a \ac{DM} halo, which could affect the accretion rate.  

\subsection{Michel accretion}

The general relativistic version of the Bondi model is known as Michel accretion \citep{Michel1972}.
Instead of describing gravity by means of a potential, it is now understood as the effect
of a curved spacetime described by the metric $g_{\mu\nu}$, which is a solution to the Einstein equations.
For a static spherically-symmetric spacetime described by the coordinates $x^\mu = (t,r,\theta,\phi)$, the line element
\begin{equation}
	ds^2 = g_{\mu \nu} dx^{\mu} dx^{\nu}\,.
\end{equation}
can be written in a general way as
\begin{equation}
	ds^2 = g_{tt}(r) dt^2 + g_{rr}(r) dr^2 + r^2 d\Omega^2 
\end{equation}
where $d\Omega^2 = d\theta^2 + \sin{\theta}d\phi^2$.

The equations of relativistic hydrodynamics consist of the conservation of mass and energy-momentum (that is, the divergence-free nature of the mass current and the stress-energy tensor), whose covariant expressions are
\begin{align}
    \label{eq:mass-conservation}
    \nabla_\mu(\rho u^\mu) &= 0\,, \\
    \label{eq:em-conservation}
	\nabla_\mu T^{\mu\nu} &= 0\,,
\end{align}
where $\nabla_\mu$ is the covariant derivative associated to the metric $g_{\mu \nu}$, $\rho$ is the fluid's rest-frame density, and $u^\mu$ is the fluid's 4-velocity.

The stress-energy tensor of a perfect fluid can be expressed as
\begin{equation}
	T^{\mu \nu} = \rho h u^\mu u^\nu + p g^{\mu \nu}
\end{equation}
where $h$ is the specific relativistic enthalpy, $p$ is the pressure in the fluid frame, and $g^{\mu \nu}$ denote the components of the inverse of the spacetime metric.

Under the assumptions of stationarity and spherical symmetry, equations \eqref{eq:mass-conservation} and \eqref{eq:em-conservation} can be manipulated to obtain the general-relativistic analogue of equations \eqref{eq:mdot-newtonian}
and \eqref{eq:bernoulli-newtonian}, namely.

\begin{align}
    \label{eq:mdot-gr}
	4 \pi \sqrt{-g_{tt}g_{rr}}r^2 u^r &= \dot{M} \,, \\
    \label{eq:bernoulli-gr}
    -h u_t &= E \,,
\end{align}
Where again $\dot{M}$ and $E$ are constants.
After assuming an equation of state and expressing the enthalpy in terms of the sound speed $c_s$,
the two algebraic equations \eqref{eq:mdot-gr} and \eqref{eq:bernoulli-gr} can be solved simultaneously
for the fluid 3-velocity $v^r = u^r/\Gamma$ (where $\Gamma$ is the Lorentz factor) and $c_s$.
This in turn allows to find the rest of the fluid variables.
More details on the solution of the Michel problem can be found, for instance,
in \citet{rezzolla_relativistic_2013} for the case of a Schwarzschild black hole,
and \cite{Meliani2016} for a general spherically symmetric metric.

In the next section, we will specialize on the particular case of the metric we use to
describe a black hole surrounded by a dark matter halo.





\begin{figure}
	
	\centering
	\makebox[\linewidth]{%
		\begin{tabular}{cc}
			\includegraphics[width=0.9\linewidth]{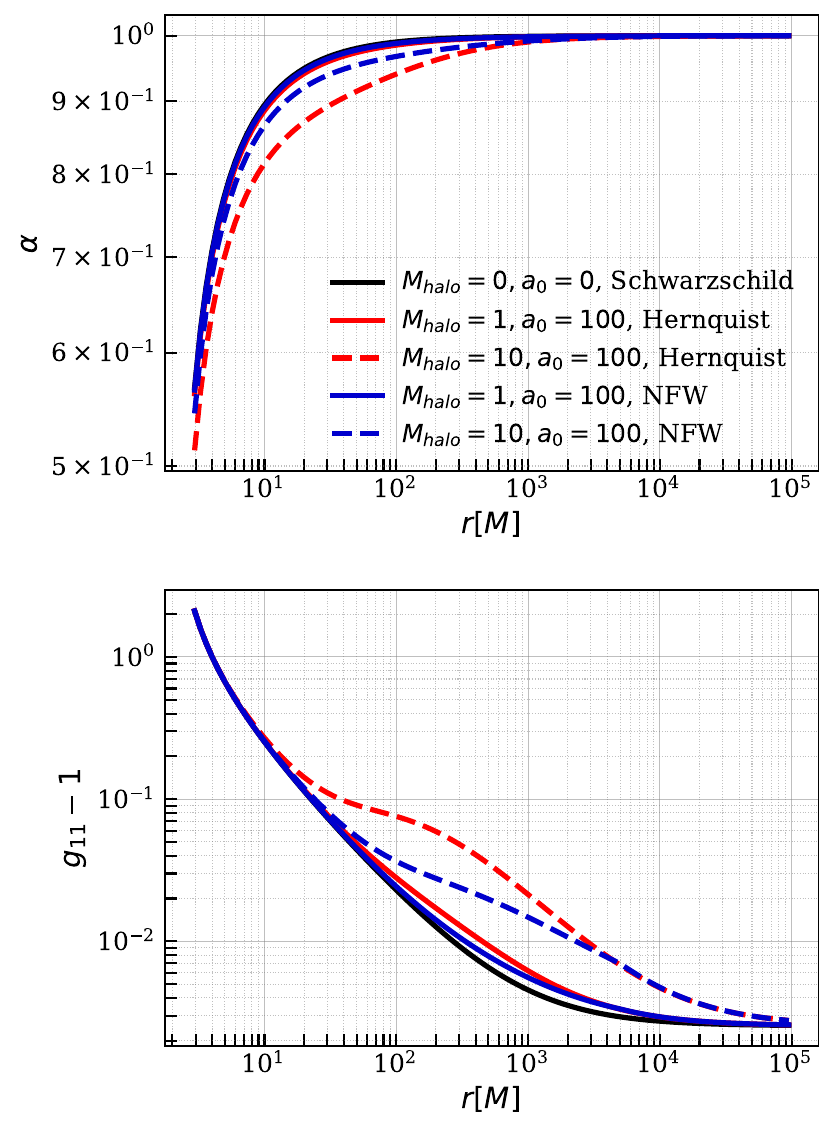}
		\end{tabular}%
	}
	\centering  
	\caption{ Metric components $g_{11}$ and $\alpha$ as functions of the radius r for the different models of black holes, Schwarzschild metric is shown with the solid blue curves. Hernquist and \ac{NFW} metrics are also shown with the blue and red curves respectively (solid and dashed line).}
	\label{metric}
	
\end{figure}

\begin{figure*}
	
	\includegraphics[width=0.9\linewidth]{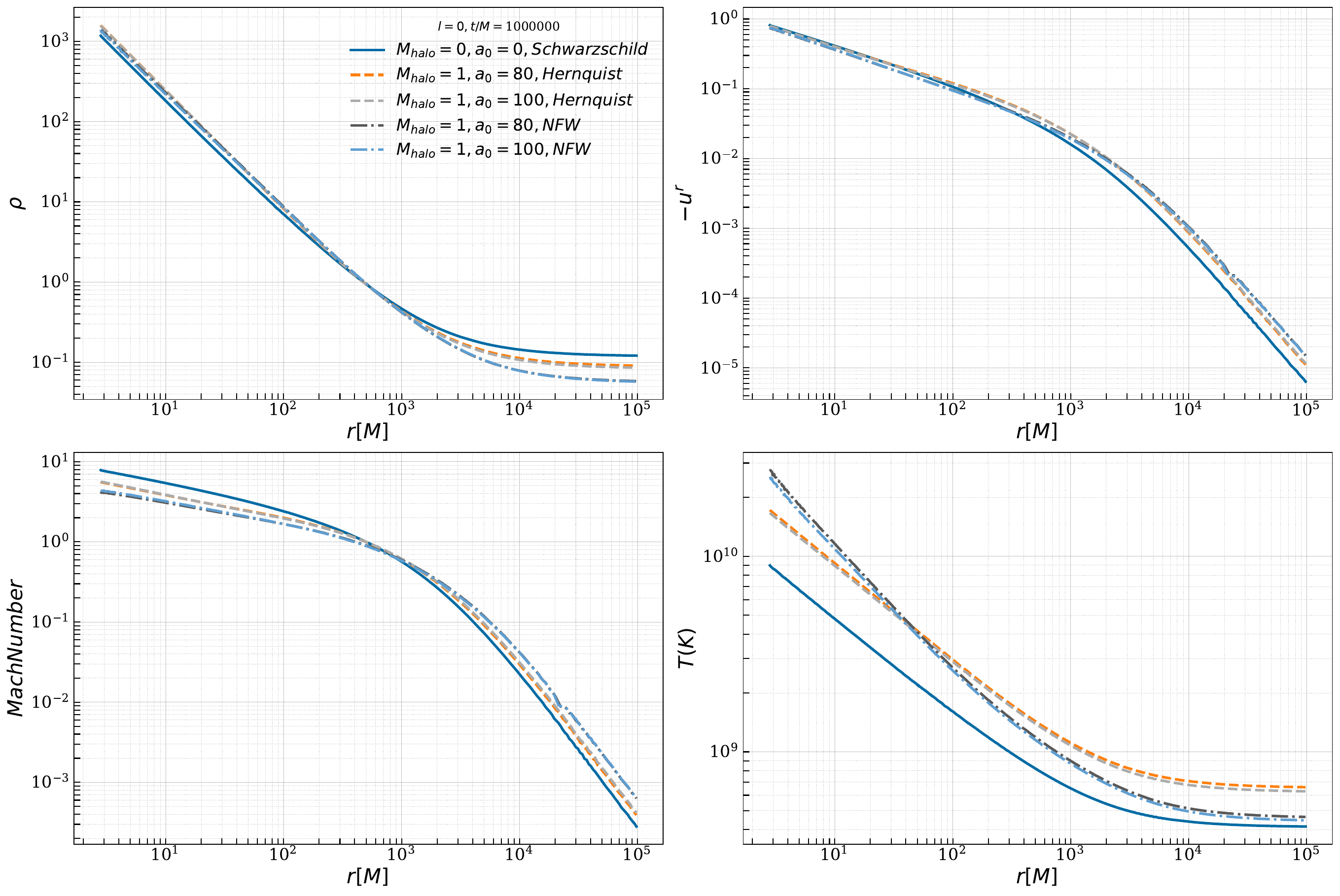}
	\centering  
	\caption{The radial component of spherical accretion parameters as a function of radius $r[M]$ in the presence of a \ac{DM} halo. These 1D simulations are shown at \( t = 1,000,000 \) for different coordinates (Schwarzschild, Hernquist, \ac{NFW}), with the outer edge of the flow set at \( r = 10^5 M \). The solid blue line represents the Schwarzschild case without a \ac{DM} halo, while the dashed and dot-dashed lines correspond to cases with a \ac{DM} halo modeled by the Hernquist and \ac{NFW} profiles, respectively. The mass of the \ac{DM} halo in these simulations equals the mass of the black hole, \( M_{\text{halo}} = M_{\text{BH}} \).}
	\label{1D_m1}
	
\end{figure*}

\section{Spacetime geometry}\label{sec.3}

We implement the newly introduced metrics to investigate the impacts of dark matter on black hole accretion.

To account for the effect of the \ac{DM} halo on the spacetime geometry,
we construct a metric that is consistent with the mass function
\begin{equation}
    \label{eq:mass-function-definition}
    m(r) = 4\pi \int_0^r \rho(r') r'^2 dr'\,,
\end{equation}
where $\rho$ is the matter density including the halo and the central \ac{SMBH}.
Similarly to \citet{2022PhRvD.105f1501C}, we do not find $m(r)$
but take it as given in such a way that is consistent with the \ac{DM} profiles
at large scales.
For simplicity, the system is assumed to be spherically symmetric.
Under such assumptions, a general static solution of the Einstein equations
can be written in the form
\begin{equation}\label{space}
	ds^2 = -f(r) dt^2 + \frac{dr^2}{1 - 2m(r)/r} + r^2 d\Omega^2 \,,
\end{equation}
where the radial function $f(r)$ is related to the mass function
by
\begin{equation}
	\label{eq:lnf_integral}
	\ln f(r) = - \int_r^\infty \frac{2m(r')}{r'(r' - 2m(r'))} dr' \,.
\end{equation}
The integral can be calculated numerically to keep
the method is flexible for any mass function,
that will correspond to a specific dark matter
profile.
To this purpose, it is convenient to work in compactified coordinates
to evaluate the integrand at infinity.
Our compactified coordinate $x$ is defined as
\begin{equation}
	x \coloneqq \frac{r}{r + 1}
\end{equation}
and the inverse transformation is
\begin{equation}
	r = \frac{x}{1 - x} \,.
\end{equation}
Using these coordinates, the integral \eqref{eq:lnf_integral} becomes
\begin{equation}
	\label{eq:lnf_integral_x}
	\ln f(r) = - \int_{x(r)}^1 \frac{2m(x')}{x'[x' - 2m(x')(1 - x')]} dx' \,,
\end{equation}
which can be integrated numerically for a known $m(x)$.
In particular, to obtain the spacetimes used in this work, we performed
the integration using Simpson's rule.
We find that for the cases considered in this work, 
a small number of points ($\sim40$)
is sufficient to achieve convergence.
In the next Sections, we will give more details on the specific density
profiles that we studied.

\subsection{Halo Density Profiles}

A wide class of density distribution of a galactic halo in astrophysics can be written as a double power law \citep{2006AJ....132.2685M, Hernquist1990GalaxyModel, Jaffe1983GalaxyModel}

\begin{equation}\label{rho_general}
	\rho_r = \frac{\rho_0}{(r/r_0)^\gamma[1+ (r/r_0)^\alpha]^{\beta - \gamma/\alpha}}
\end{equation}

If $r << r_0$, $ \rho \sim r^{-\gamma}$ and if $r>>r_0 $, $ \rho \sim r^{-\beta}$. 
The parameter $\alpha$ controls the sharpness of the break. In order for the mass to be finite, it requires $\gamma < 3$ and $\beta > 3$. 
Table \ref{table1} shows a list of double power-law density profiles frequently used in astronomy.

\begin{figure*}
	
	\centering
	\includegraphics[width=0.9\linewidth]{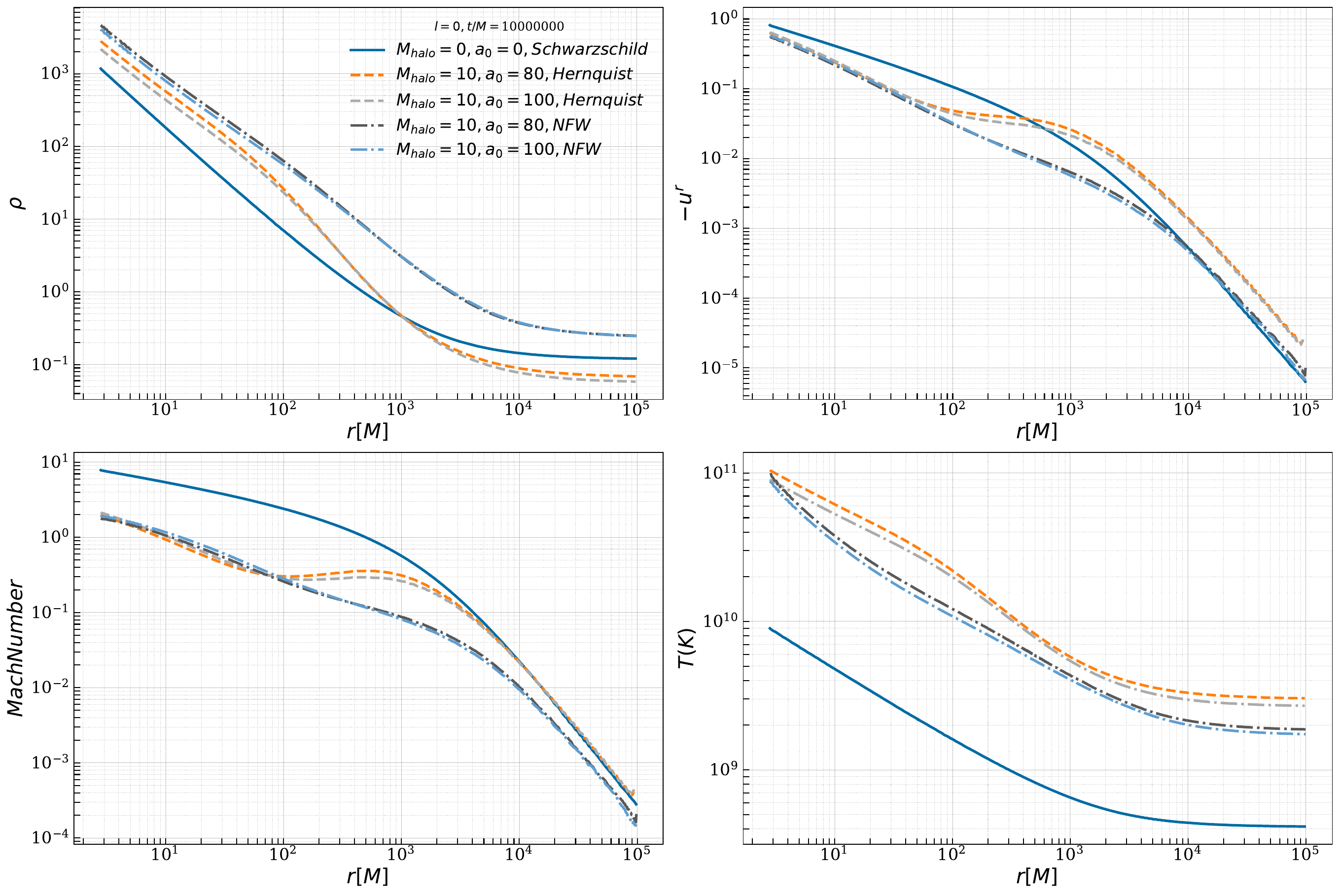} 
	\centering  
	\caption{Similar to Fig. \ref{1D_m1}, but is assumed the mass of the halo to be $10$ times the black hole mass \( M_{\text{halo}} = 10 M_{\text{BH}} \).}
	\label{1D_m10}
	
\end{figure*}

The Hernquist profile models mass distributions in elliptical galaxies, with a central concentration falling off as \( r^{-1} \) at small radii and as \( r^{-4} \) at large radii. In contrast, the \ac{NFW} profile is commonly used for \ac{DM} halos, with a central density decaying as \( r^{-1} \) at small radii and \( r^{-3} \) at large radii. Using both profiles can help us understand \ac{DM}'s role in galactic centers and provide insights into how dark matter distribution influences galactic structure formation.
\\

\subsubsection{Hernquist profile}
The Hernquist profile \cite{Hernquist1990GalaxyModel} $\alpha = 1, \beta = 4, \gamma = 1 $ is applied for modeling the Sérsic profiles observed in bulges and elliptical galaxies \cite{Konoplya2022BlackHoleHalo}. As a first step, we use the mass function inspired by the Hernquist profile that appears in \citet{2022PhRvD.105f1501C}. The \ac{BH}+\ac{DM} halo system can be modeled by the mass function
\begin{equation}
	m(r) = M_{\rm BH} + \frac{Mr^2}{(a_0 + r)^2}
	\left(1 - \frac{2 M_{\rm BH}}{r}\right)^2 \,,
\end{equation}
where $M_{\rm BH}$ is the black hole mass, $M$ is the halo mass,
and $a_0$ is the halo scale parameter.

In terms of the compactified coordinates,
the mass function to be inserted in \eqref{eq:lnf_integral_x} reads
\begin{equation}
	m(x) = M_{\rm BH} + M \left(\frac{x - 2 M_{\rm BH} (1 - x)}{x + a_0 (1 - x) }\right)^2 \,.
\end{equation}

\subsubsection{NFW Profile}

Although the Hernquist profile describes isolated dark-matter halos, because of the continued infall from the cosmological background, halos found in cosmological simulations are better fit by an \ac{NFW} profile. For galaxies with the largest content of dark matter, the Navarro-Frenk-White model Navarro et al. (1995, 1997) $ (\alpha = 1, \beta = 3, \gamma = 1) $ is mostly used \cite{Konoplya2022BlackHoleHalo}.

The density profile of \ac{DM} is the \ac{NFW} profile obtained from numerical simulations based on \ac{DM} and $\Lambda CDM$. Here, we use the mass function inspired by the \ac{NFW} profile that appears \cite{Konoplya2022BlackHoleHalo}.

\begin{equation}
	m(r) = \frac{r}{2}(1 - (1-\frac{2M_{BH}}{r})(1 - \eta \frac{a}{r+a} - \eta \frac{a}{r} ln\frac{a}{r+a}))
\end{equation}

where

\begin{equation}
	\mu = \frac{a}{a+s} \quad 
\end{equation}

\begin{equation}
	\eta = \frac{2Ms(a+s)}{a(s-2M_{BH})(s+(a+s)ln(a/(a+s)))}
\end{equation}

Here, $a$ represents the characteristic scale of the galactic halo, $M$ is the total mass of the galaxy, and $s$ is the radius of the galactic halo, which in our study is taken to be $ s = 50a $. 

In terms of the compactified coordinates, the mass function read

\begin{equation}
	m(x) = \frac{x}{2(1-x)}(1 - (\frac{x - 2M_{BH}(1-x)}{x})
\end{equation}
\begin{equation*}
	(1 - \eta \frac{a(1-x)}{x+a-ax} - \eta \frac{a(1-x)}{x} ln\frac{a(1-x)}{x+a(1-x)}) )
\end{equation*}

\begin{table*}[ht]
\centering
\label{tab:models}
\begin{tabular*}{\textwidth}{@{\extracolsep{\fill}}lllllllll@{}}
\toprule
\hline
Model & Simulation & Halo mass & Length scale & resolution & $ l $ & Outer Boundary & Sonic Point & $\dot{M}_\text{BH}/\dot{M}_\text{Bondi}$ \\
\midrule
Schwarzschild & 1D & $0.0$ & $0.0$ & $96$ & $0$ & $ 10^5$ & $ 500 $ & $1.0$ \\
Hernquist & 1D & $1.0$ & $100.0$ & $96$ & $0$ & $10^5$ & $ 500 $ & $1.3058$ \\
Hernquist & 1D & $10.0$  & $100.0$  & $96$ & $0$ & $10^5$ & $500$ & $1.3605$ \\
NFW & 1D & $1.0$  & $100.0$ & $96$ & $0$ & $10^5$ & $ 500 $ & $1.0675$ \\
NFW & 1D & $10.0$  & $100.0$ & $96$ & $0$ & $10^5$ & $ 500 $ & $2.3782$ \\
\hline
Schwarzschild & 2.5D & $0.0$ & $0.0$ & $104 \times 96$ & $0$ & $ 10^3$ & $500$ & $1.0$ \\
Hernquist & 2.5D & $1.0$ & $100.0$ & $104 \times 96$ & $0$ & $10^3$ & $ 500 $ & $1.2889$ \\
Hernquist & 2.5D & $1.0$ & $100.0$ & $104 \times 96$ & $3.25$ & $ 10^3$ & $ 500 $ & $0.8101$ \\
NFW & 2.5D & $1.0$ & $100.0$ & $104 \times 96$ & $0$ & $ 10^3$ & $500$ & $1.0804$ \\
NFW & 2.5D & $1.0$ & $100.0$ & $104 \times 96$ & $3.25$ & $ 10^3$ & $500$ & $2.2672$ \\
\bottomrule
\end{tabular*}
\caption{Simulation parameters and results. Columns 1 and 2 list the dark matter halo profile and simulation dimensions. Columns 3 and 4 correspond to the halo mass and its characteristic length scale, while column 5 provides the resolution. Column 6 shows the angular momentum of the accretion flow. Columns 7 and 8 indicate the outer boundary and the sonic point location, respectively. Finally, column 9 presents the mass accretion rate normalized to the Bondi accretion rate.}
\label{table}
\end{table*}

We adopt the density distribution (\ref{rho_general}), where the constant \( \rho_a \equiv \rho(a) \) determines the total mass of the galaxy (\cite{Konoplya2022BlackHoleHalo}). 

\begin{equation}\label{mass}
	M = m(s) = 4\pi \int_0^s \rho(r) dr
\end{equation}

The parameter \( s \) represents the radius of the halo, with \( s > a \gg M \). Notably, for \( \beta > 3 \), the galaxy size 
can be considered infinite, as the improper integral in Equation \ref{mass} converges as \( s \to \infty \). When \( s \) is finite, 
to ensure a finite total mass, we assume that the space beyond the galactic halo is empty, i.e., \( \rho(r > s) = 0 \) similar to \cite{Konoplya2022BlackHoleHalo}.

In Figure \ref{metric}, we plot the metric as a function of radius. Notably, the metric functions of a non-rotating black hole embedded in dark matter deviate from those of a Schwarzschild \ac{BH} in a vacuum. However, near the event horizon, these deviations remain minimal due to the strong gravitational influence of the black hole in this region. Also, the plots from both the Hernquist and NFW sections show that this deviation increases with increasing halo mass.

\begin{figure*}
	
	\centering
	\includegraphics[width=0.9\linewidth]{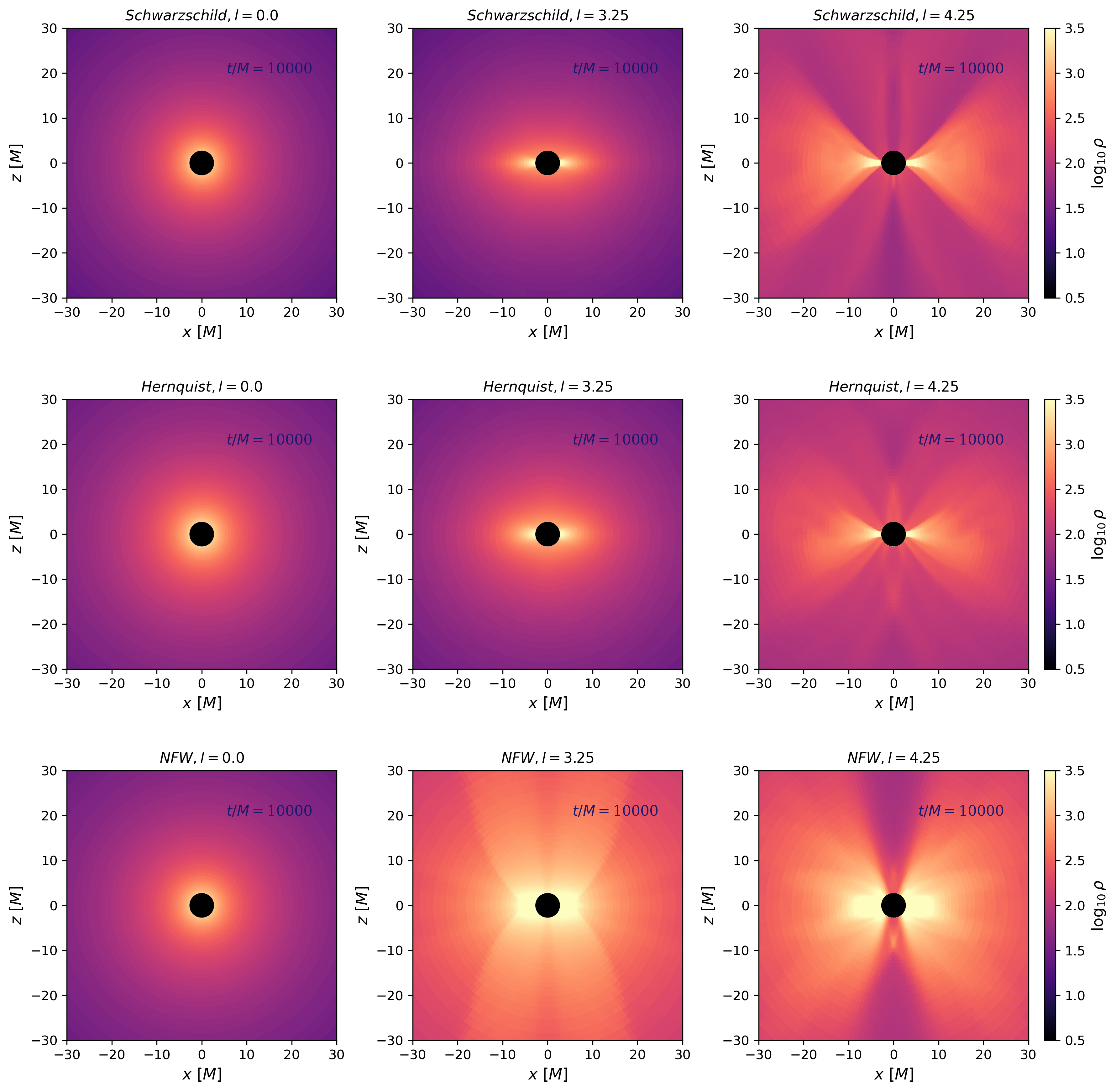} 
	\caption{Snapshots of 2.5D \ac{GRHD} simulations from the near regions to the \ac{BH} for different metrics, Schwortzshild, Herenquist, \ac{NFW} (up to down) at time $t/M = 10000 $. The colors show the logarithm of gas density $\rho$. The left to right show (zero angular momentum, $l = 0$, $l = 3.25$, $l = 4.25$ ). In all simulations, quasi-spherical accretion flow structures form around the \acp{BH}.}
	\label{density}
\end{figure*}

\section{Simulation Setup}\label{Setup}
To consider the effect of galactic halo on the dynamics of the transonic spherical accretion flow, we performed a set of (1D-2.5D) ideal \ac{GRHD} simulations by using the \ac{GRMHD} code \ac{BHAC} \citep{Porth2017BHAccretionCode, Olivares2019BHAccretionCode}.
We used a spherical polar grid in Schwarzschild coordinates with logarithmic spacing for better resolution at small radii.
Our simulations adopt units where $ G = c = 1 $, setting the gravitational radius $r_g = M$, with $M$ the black hole mass.
There is only one sonic point $r_\mathrm{s}$ in this kind of solution and we fix it at $500M$. We performed 1D and 2.5D simulations, with results discussed in sections \ref{sec:1Dresults} and \ref{sec:2.5Dresults}, respectively. For 1D simulations, we assume zero angular momentum and set the outer boundary at $ r = 10^5 M $. In 2.5D simulations, small amounts of angular momentum are incorporated, and the outer boundary is located at $r = 10^3$. To perform the 1D simulations, we consider an effective resolution ($96$). For each halo mass ($M_h = 1$ and $M_h = 100$), simulations are performed for two values of the length scale: $80M$ and $100M$.
The 2D simulations were performed using different coordinate systems (see Section \ref{sec.3}) with a resolution of \( N_r \times N_\theta = 104 \times 96 \) over a computational domain defined by \( r \in [2.8 M, 10^3 M] \) and \( \theta \in [0, \pi] \). For 2.5D simulations, we set the halo mass to $1$ and the length scale to $100$. Simulations are performed up to $t/M = 10^6 $ and $5\times10^3$ for 1D and 2.5D, respectively. The details of the different runs are summarized in Table \ref{table}. The fluid follows an ideal equation of state with \( \hat{\gamma} = 4/3 \). 

\begin{figure*}
	\centering
	\includegraphics[width=0.9\linewidth]{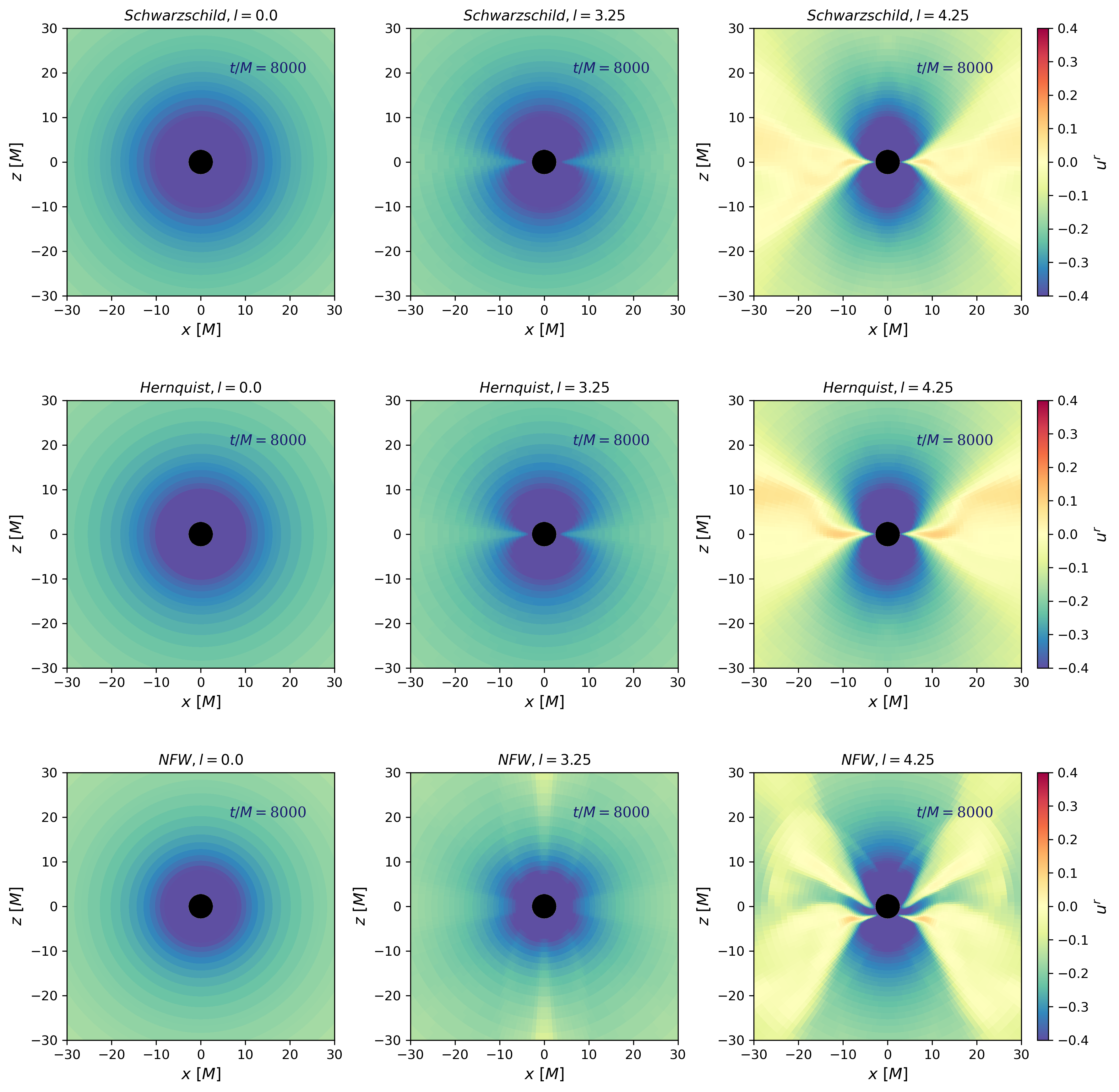}
	\caption{Snapshots of 2D \ac{GRHD} simulations of accretion and outflow the near regions to the \ac{BH} for different metrics, Schwarzschild, Hernquist, \ac{NFW} (up to down) at time $t/M = 8000 $. The colors show the gas velocity $ur$. The left to right show (zero angular momentum, $l = 0$, $l = 3.25$, $l = 4.25$ ). In all simulations, quasi-spherical accretion flow structures form around the \acp{BH}.}
	\label{u}
\end{figure*}

\section{Results}\label{result}

\begin{figure*}
	\centering
	\includegraphics[width=0.8\linewidth]{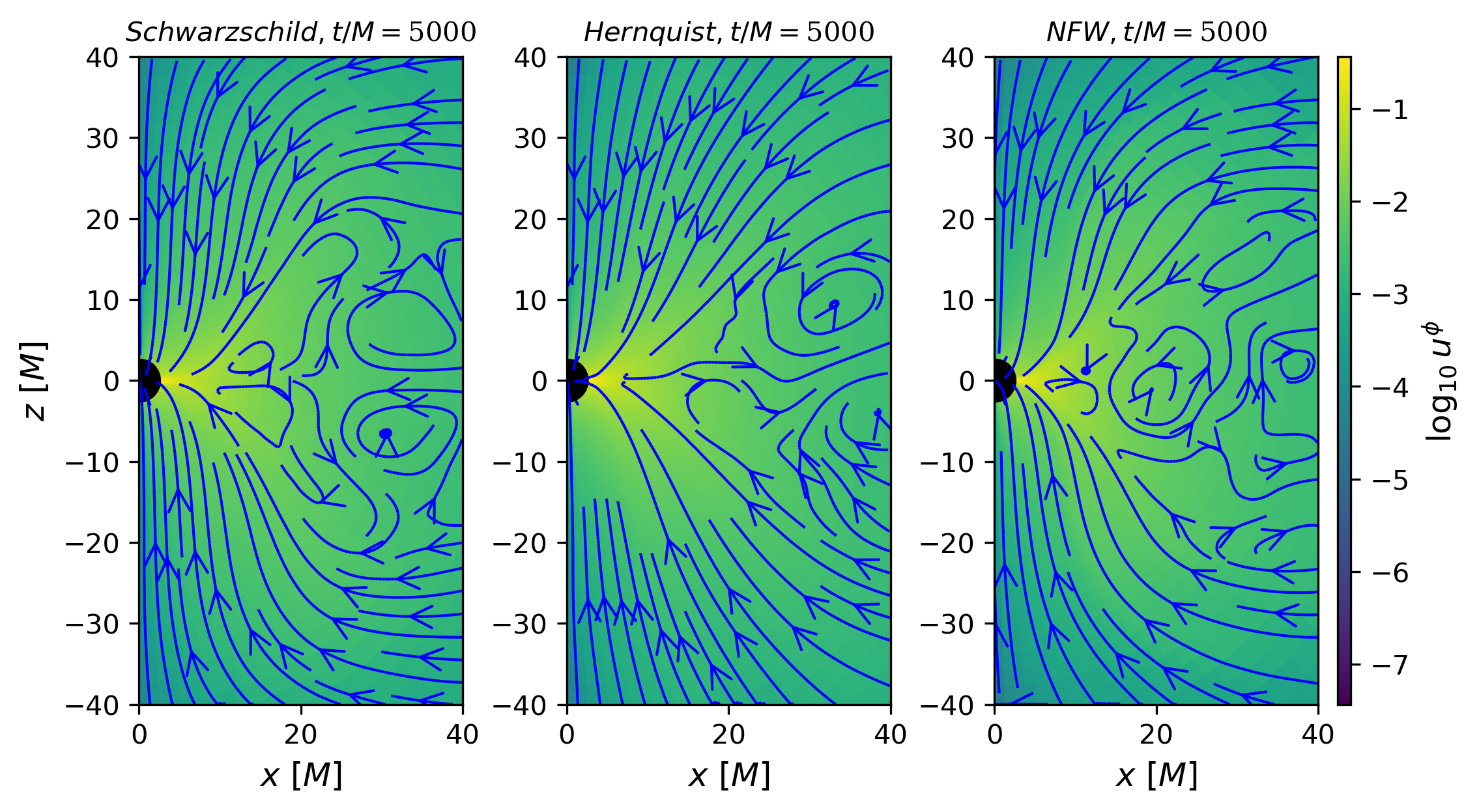}
	\caption{Velocity streamlines in the x-z plane, show components of velocity $u_x$ and $u_z$, for Schwarzschild, Hernquist, and \ac{NFW} profiles at $t/M=5000$, with angular momentum  $l=4.25$. The color bar indicates the logarithmic intensity of $u_{\phi}$ (azimuthal velocity). Each panel displays a gravitational profile's unique flow patterns.}
	\label{u3}
\end{figure*}

\subsection{Test evolutions 1D }
\label{sec:1Dresults}
In this part, we have carried out 1D simulations to explore dynamics in accretion flow with zero angular momentum. 
Figures \ref{1D_m1} and \ref{1D_m10} illustrate the impact of a dark matter halo on the dynamics of accretion flows around \acp{SMBH}, comparing these flows to those around a Schwarzschild \ac{BH} in a vacuum. In Figure \ref{1D_m1}, the halo's mass is equal to the black hole's mass, while in Figure \ref{1D_m10}, the halo’s mass is ten times larger than the black hole's.
The solid blue line represents the Schwarzschild profile which represents a static, non-rotating black hole in a vacuum, without the influence of additional mass like dark matter. The (orange, gray) dashed lines correspond to the Hernquist dark matter profile, while the (black, blue) dot-dashed lines represent the \ac{NFW} profile. As the length scale increases, dark matter distribution will extend further, potentially affecting the outer accretion regions more significantly. These values apply to both Figures \ref{1D_m1} and \ref{1D_m10}.

The top-left panel illustrates the density profile as a function of radius, showing a density increase near the \ac{BH} in the presence of a dark matter halo. The dark matter halo strengthens the gravitational potential near the \ac{BH} and within the halo, causing a higher density in this region. The solid blue line shows the simulation in the Schwarzshild metric. The difference between the Hernquist and \ac{NFW} profiles is in the steepness of their density distribution. The top-right panel presents the radial velocity, which decreases inside the dark matter halo region as the gravitational effect of the halo weakens the inflow and increases outside it, where the gravitational effect of the halo is weaker.  The bottom-left panel shows the Mach number which is the speed of the flow relative to the local sound speed, and the bottom-right panel depicts temperature. Temperature also increases due to the dark matter halo. One of the reasons for the increase in temperature in the presence of the halo is likely due to the additional gravitational potential energy from the dark matter, which heats the accreting matter. Notably, the deviations in these profiles are more pronounced when the halo mass is higher, as seen in Figure \ref{1D_m10}.
One of the striking results of our simulations is that the structure of flow inside and outside of the radius of the halo is different, so a huge halo can affect a wide region of accretion flow and increase the mass accretion rate significantly. In particular, we find an increase in the accretion rate by up to over 100\% when dark matter halos are present, especially when the halo mass is large.
We have restricted our study to a small halo which is strictly aligned with the early universe and high redshift (see the section \ref{com} for more details).

\subsection{2D simulation}
\label{sec:2.5Dresults}
In astrophysical accretion flows, the presence of angular momentum is anticipated, and it undoubtedly influences various flow properties, including the mass accretion rate. The behavior of Bondi-like accretion flows with low angular momentum has also been extensively studied with large-scale hydrodynamic simulations (e.g. \citep{Olivares2023GRHydroSimulations, 2024ApJ...967....4D, 2003ApJ...592..767P, 2017MNRAS.472.4327S}). We are primarily focused on the lower limits of angular momentum and conclude that the behavior of accretion flow differs significantly from that of high and zero angular momentum flows. Strikingly, the mass accretion rate in rotating accretion flows with low angular momentum, encompassed by a dark matter halo, remains unexplored. It is plausible to expect that if the rotational support is insignificant at the Bondi radius, the impact of angular momentum on the mass accretion rate would be minimal. 

In this section, we investigated the effect of both profiles of dark matter halo on accretion flow by evolving an accretion flow with angular momentum ($l = 3.25, 4.25$) in 2.5D, as illustrated in Figure \ref{density}, \ref{u}, \ref{u3}. As we expected, in the high angular momentum simulation ($l = 4.25$), we observe noticeable shock and turbulence in the region close to the event horizon. 

Figure \ref{density} shows snapshots of logarithmic density maps for all simulations at $t = 10 000 M$ on the $x-z$ plane. The snapshot at $t=10000M$ represents a steady-state phase that provides the same conditions for comparison. Panels are organized as increasing angular momentum from left to right and different metrics from top to bottom (Schwarzschild, Hernquist, \ac{NFW}). The snapshots indicate denser accretion structures near the \ac{BH} for the Hernquist and \ac{NFW} profiles compared to a less dense pattern in the Schwarzschild metric. The density maps in the Schwarzschild panels indicate symmetry and uniformity, especially at $l=0$, showing the simplicity of the accretion flow in Schwarzschild profiles. We can see clear differences in simulations with higher angular momentum. Movie of simulation is available.

Figure \ref{u} displays snapshots of radial velocity in the x-z plane. The rows, from top to bottom, depict three different metrics: the first row represents the Schwarzschild metric without dark matter, the second row includes the Hernquist profile, and the third row corresponds to the \ac{NFW} profile of the dark matter halo. In these panels, angular momentum values increase from left to right, with $l = 0$, $l = 3.25$, and $l = 4.25 $, respectively.

In these snapshots, blue regions indicate gas accreting into the \ac{BH}, while red regions show outflows of matter being removed from the accretion flow. The results demonstrate that outflows become stronger as angular momentum increases, in all metrics. For $l = 0 $, accretion occurs without any outflow. The presence of a dark matter halo enhances the strength of the outflow, as observable in this figure. Movies of simulations are available.

Figure \ref{u3} shows a streamline of components of velocity $u_x$ and $u_z$, while the background color map presents the logarithm of the intensity of velocity in the direction of $\phi$ that is the azimuthal velocity ($u_{\phi}$). We run all three simulations related to Schwarzschild, Hernquist, and \ac{NFW} profiles at $t/M = 5000$. In this case, we have a high angular momentum of $l = 4.25$. The streamlines show how gas moves around the \ac{SMBH} in each simulation. In the Schwarzschild profile, the flow is approximately radial with minimal rotational structure. In the Hernquist profile, we can see the onset of circularized motion, which is indicative of the effects of angular momentum. The \ac{NFW} profile exhibits more complex structures, which may be related to a higher accretion rate. Movies of their evolution are available.
The mass accretion rate for the \ac{NFW} profile is higher than that of the Hernquist profile. 

The simulations with the impact of dark matter halo are summarized in Table \ref{table}. We therefore performed many simulations with different values of the halo mass and characteristic length scale at both large and small scales. In column 9 of Table \ref{table}, we present the mass accretion rate (in units of the Bondi accretion rate). Table \ref{table} includes two simulation models: zero angular momentum and low angular momentum. For the zero angular momentum 1D simulations, the outer radial boundary is set at $ 10^5 [M]$. For the low angular momentum 2.5D simulations, the outer boundary is $ 10^3 [M]$. We expect the mass accretion rate for an accretion flow with angular momentum to be lower than in the case without angular momentum. However, since our simulation with the \ac{NFW} profile is not fully quasi-stationary, the value of $2.2672$ in Table \ref{table} is not entirely physical. In contrast, the Hernquist profile exhibits better stability, indicating that it reaches a quasi-stationary state faster than the \ac{NFW} profile.

\begin{figure*}
	\centering
	\includegraphics[width=1.0\linewidth]{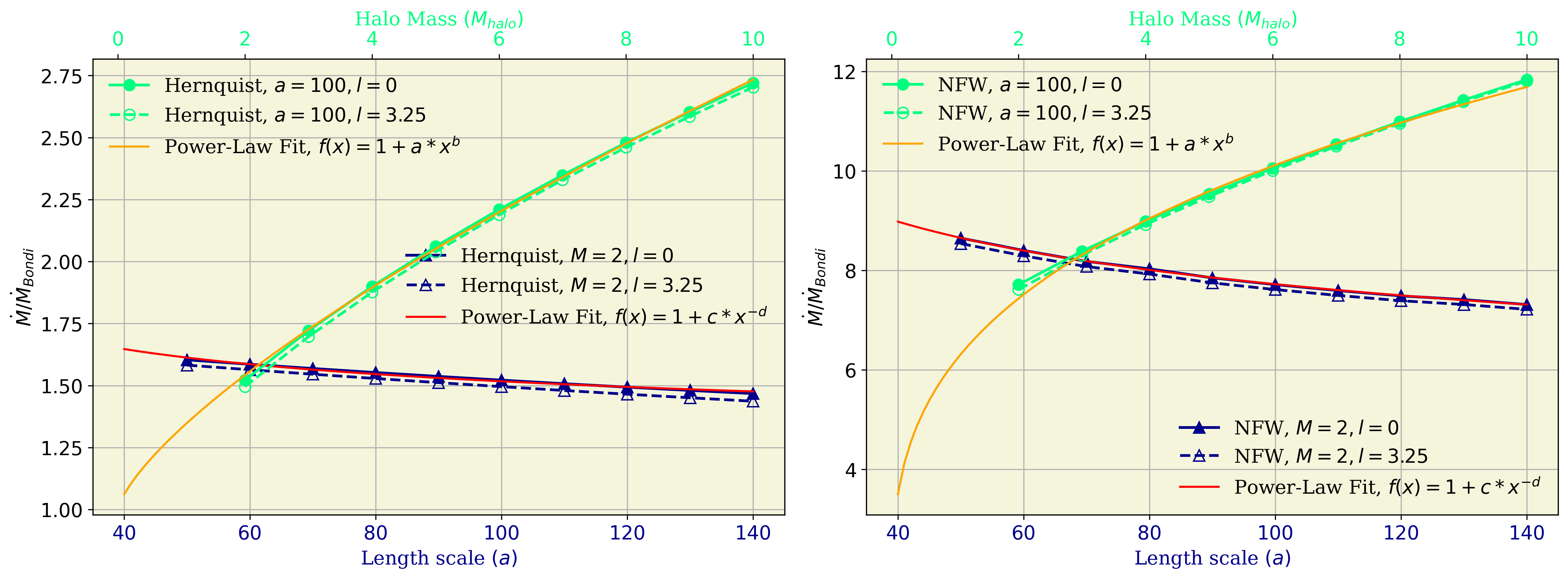} \\
	\caption{Mass accretion rate normalized by Bondi accretion rate, as a function of length scale (top x-axis, green) and halo mass (lower x-axis, blue) for Hernquist (left panel) and \ac{NFW} (right panel) dark matter profiles. Solid lines represent zero angular momentum ( $l = 0$ ), and dashed lines indicate ( $l = 3.25$ ). The orange and red solid lines are power-law fit for these data. The coefficients of the Power-law function are provided in Table \ref{tab:coefficients}.
    }
	\label{dot}
\end{figure*}

Figure \ref{dot} illustrates the variation in mass accretion rate as influenced by the length scale and halo mass for Hernquist and \ac{NFW} profiles. The left panel corresponds to the Hernquist profile, and the right panel corresponds to the \ac{NFW} profile. In these simulations, we explored angular momentum values of ( $ l = 0 $) (solid lines) and ( $l = 3.25$ )  (dashed lines). In both panels, the blue curves depict variations with respect to the length scale (lower x-axis), while the red curves represent changes as a function of the halo mass (upper x-axis). and It is evident from the figure that adding angular momentum reduces mass accretion, as previously demonstrated in simulations and analytical studies (\cite{Ranjbar2023}). The simulations were performed at ( $t/M = 100,000$ ). Notably, the \ac{NFW} profile exhibits a higher mass accretion rate compared to the Hernquist profile, a trend consistent across both zero and non-zero angular momentum scenarios. This illustrates the stronger gravitational binding of the \ac{NFW} profile. 
We performed power-law fits to model the behavior of the mass accretion rate for both the Hernquist and ac{NFW} profiles. The results of the fits are presented for different configurations (\(l = 0\) and \(l = 3.25\)), and the coefficients of the fitted power-law are provided for each case in Table \ref{tab:coefficients}.

These power-law fits accurately represent the behavior of the mass accretion rate for the Hernquist and \ac{NFW} profiles. The red curves capture the general trends of the mass accretion rate with respect to the length scale, while the orange curves account for variations with halo mass. This analysis provides deeper insight into how the accretion rate changes with key parameters, aiding in the interpretation of physical phenomena in cosmology.

\begin{table}[h]
    \centering
    \begin{tabular}{|p{3cm}|p{1cm}|p{1cm}|p{1cm}|p{1cm}| }
    \hline 
        \textbf{Profile (Case)} & \( \textbf{a} \) & \( \textbf{b} \) & \( \textbf{c} \) & \( \textbf{d} \) \\ 
        \hline \hline
        Hernquist (\( l = 0 \))   & 0.3287  & 0.7221 & 1.5957  & 0.2447  \\ 
        Hernquist (\( l = 3.25 \)) & 0.3109  & 0.7426 & 1.733  & 0.2742  \\ 
        \hline \hline 
        NFW (\( l = 0 \))         & 5.2724  & 0.3075  & 15.9134 & 0.1871 \\ 
        NFW (\( l = 3.25 \))       & 5.1724  & 0.3151 & 15.7361  & 0.1880 \\ 
      \hline \hline 
    \end{tabular}
    \caption{Power-law coefficients for Hernquist and NFW profiles across different cases of \( l \) as shown in Fig. \ref{dot}.}
    \label{tab:coefficients}
\end{table}

\subsection{Relevance to observed systems}\label{com}
Even though the parameters of the \ac{BH}-halo system considered in our simulations are not representative of nearby galaxies, they more closely describe the mass relation between \ac{DM} halos and \acp{SMBH} at high redshift in the early universe. It is believed that the relationship between black hole mass $M_{BH}$ and the mass of the dark halos $M_{h}$ in which it grows gives different insights into coevolution because it directly constrains the \ac{SMBH} growth efficiency in halos. \acp{SMBH} in luminous \acp{QSO} expand much more rapidly than host halos, owing to gas accretion, although we cannot rule out the possibility that undetected \acp{SMBH} may have a less local $M_{BH}/ M_{h}$ ratio. 
Also, preliminary evidence suggests that smaller halos that $5 \times 10^{11} M_{\odot}$ are becoming less efficient at forming \acp{SMBH}, possibly to the point of being unable to do so. Our simulation limit is in the special range of $M_{BH}/ M_{h}$ that appears in some high redshifts \acp{QSO} such as J1044 $-$ 0125 \cite{Shimasaku2019}. According to some observations, the $M_{BH}/M_{h}$ increases with redshift. Furthermore, the semi-analytical model of \cite{Shirakata2019ν2GC} shows that $M_{BH}/M_{h}$ increases with redshift. Some hydrodynamical simulations show that $M_h \sim 10^{12} M_{\odot} $ halos can have an \ac{SMBH} as massive as $ \sim 10^9 M_{\odot}$ (e.g.\cite{Tenneti2019}), but based on only several examples. The James Webb Space Telescope (JWST) has provided results for studying and understanding high redshift quasars $z \geq 6$ powered by rapidly accreting supermassive black holes with masses reaching up to $\sim 10^9 $. For instance, the quasar J1120 $+$ 0641 is driven by an exceptionally massive black hole, which exhibits the highest reported black hole-to-stellar mass ratio ($M_{BH}/M_*$) of approximately $0.54$ (\cite{Marshall2024QuasarHost}). It is a representative example of high redshift quasars with supermassive black holes that we discussed in this section. Our results can be used to calibrate the efficiency of \ac{SMBH} growth in the early cosmic epoch. This means our simulation can be applicable to a galaxy with high redshift and in the early universe. 
Two of the most well-known massive black holes, \sgra and \m{87*}, possess halo masses of $1.91 \times 10^{11} M_{\odot}$ and $1.3 \times 10^{14} M_{\odot}$, respectively, with characteristic length scales of $20kpc$ and 9$91.2 kpc$ (\citep{Posti2019MWDarkMatterHalo}, \citep{DeLaurentis2022M87}). The mass difference between the central black hole and the halo is on the order of $10^4$, so simulating these systems requires high-performance computing and significant resources. A detailed investigation of these objects, along with a comparison between simulation results and observational data, will be reserved for future studies.

\subsection{Thermal Bremsstrahlung emissivity} 
We aim to incorporate an observational perspective into our simulation, highlighting the presence of dark matter and quantifying its impact by comparing Bremsstrahlung's emission with previous works. Based on Chandra's observations of massive black holes in galaxies (\cite{Baganoff2003}), we can interpret the results in this paper, which helps to validate our simulations. In such systems, the black hole may accrete the hot interstellar medium (ISM) surrounding the galaxy. Bremsstrahlung (free-free radiation) is radiation emitted from charges accelerated in the Coulomb field of another charge. Energy is emitted at the expense of kinetic energy, thus free-free radiation can cool the plasma. In most cases,  the Bremsstrahlung radiation is produced at large radii, where ions' temperatures are not significantly different from electron temperatures, so we consider one temperature the Bremsstrahlung radiation. Bremsstrahlung is the main cooling process for high T ($>10^7$ K) plasma. Integrating over the whole spectrum has only that the non-relativistic thermal bremsstrahlung per unit area, $e_{ff}$ is given by
\begin{equation}
	e_{ff} = 1.4 \times 10^{-27} T^{1/2} Z n_e n_i g_B \quad 
\end{equation}
\begin{equation*}
	= 5.54 \times 10^{-15} Z^2 \times (\frac{m_i}{m_p})^{1/2} (\frac{n_i \rho}{10^6 })^2 (\frac{p}{\rho})^{1/2}
\end{equation*}

when relativistic corrections are not considered. The unit in CGS, (erg $s^{-1}$ $cm^{-3} $, radiated power per cubic cm). $g_B$ is the frequency of the velocity averaged Gaunt factor, which is in the range of $1.1$ to $1.5$.

\begin{equation}
	L_{BR} = 2.41 \times 10^{38} (\frac{M}{4.15 \times 10^6 M_{\odot}})^3 \int e_{ff} \Gamma \sqrt{\gamma} d^3 x
\end{equation}
The unit in CGS, erg$/s$. $\Gamma$ is the Lorentz factor. Similarly to \cite{Olivares2023GRHydroSimulations}, we have chosen the parameters for monoatomic hydrogen. Since Bondi and ADAF's 'classical' models assume the X-ray emission from \sgra is dominated by thermal bremsstrahlung emission from electrons in a hot, optically thin accretion flow \cite{Baganoff2003}. We have calculated the Bremsstrahlung emission for \sgra\ using different density profiles. For \sgra with black hole mass $ 4.15 \times 10^6 M_{\odot}$, luminosity is equal to  $7.63 \times 10^{40}$ erg s$^{-1}$ for Hernquist profile and $ 2.92 \times 10^{40}$ erg $s^{-1}$ for \ac{NFW} profiles. 
These values are scaled by a factor of $ ({n_i}/{10^6})^2 $.
These values are not realistic for \sgra; however, they serve as an illustrative example. This demonstrates that the \ac{DM} profile can influence observational properties, as evidenced by a more than twofold increase in Bremsstrahlung luminosity for the considered halo mass.

\section{Discussion and Conclusions}\label{sec.5}

 A tight correlation exists between the mass of a black hole and its host \ac{DM} halo, indicating a connection between the two in terms of formation and growth. Also, \ac{AGN} feedback extends to larger scales, regulating the growth of the galaxy, black hole, and halo together.
In this study, we performed 1D-2.5D \ac{GRHD} simulations to model the spherical accretion flow around a black hole surrounded by a \ac{DM} halo. The presence of \ac{DM} significantly alters the structure and dynamics of the accretion flow, deviating from the characteristics observed in a vacuum scenario with a Schwarzschild black hole. These accretion scenarios were evaluated using several cases, including accretion flow with zero and low angular momentum, for several quantities, and different \ac{DM} haloes used to analyze their properties. The altered flow structures and enhanced accretion rates suggest that \ac{DM} not only affects the large-scale dynamics of galaxies but also has a profound impact on the region near the supermassive black holes. 

The gravitational influences of a \ac{DM} halo on spherical accretion flows are analyzed by comparing the results with the exact general relativistic Bondi solution. Two independent models of \ac{DM} halos were considered to evaluate their gravitational effects, aiming to reproduce general relativity Bondi accretion under more realistic conditions. These findings underscore the significant role of \ac{DM} in modifying the physical properties of accretion flows. The main findings of this study can be summarized as follows:

\begin{itemize}
	
	\item Within the \ac{DM} halo region, the radial velocity of the accretion flow decreases, whereas it increases outside the halo. The presence of a \ac{DM} halo also leads to an increase in density and pressure. \\
	
	\item The temperature of the accretion flow increases by more than $100\%$ in the presence of a \ac{DM} halo for both the Hernquist and \ac{NFW} profiles. ($ \sim 10^{11}$ K near the event horizon)\\
	
	\item The size and mass of the halo significantly affect the accretion flow structure. Larger and more massive halos ($M_h = 10, a = 100$) induce larger changes in the dynamics and structure of the accretion flow.\\
	
	\item Simulations have shown that \ac{DM} can deviate the mass accretion rate from the Bondi accretion rate. The mass accretion rate, $\dot{M}$, increases significantly in models that include a \ac{DM} halo, with our simulation results showing a $\sim 30 \%$ increase for the Hernquist profile and over a $\sim 50 \%$ increase for the \ac{NFW} profile.  \\

        \item Conversely, a reduction in the mass accretion rate was observed for models with higher angular momentum, consistent with results from previous studies.\\
        
        \item  The gravitational effects of \ac{DM} profile or underlying space-time geometry can significantly influence the intensity of both inflow and outflow in low-angular-momentum accretion flows, impacting mass loss and the overall mass accretion rate.

\end{itemize}

These results have important implications for understanding supermassive black hole growth, as the altered accretion rates could affect the black hole mass and the co-evolution of black holes with their host galaxies. We have restricted our study to a small halo which is strictly aligned with early universe and high redshift.  

It remains unclear whether this change carried over to the ultra Bondi radius regime of most astrophysical interest. Therefore, in future studies, it would be interesting to explore the impact of a huge \ac{DM} halo around the Bondi radius, as well as an interstellar gas, relative to the outer boundary.
These model problems could help in understanding the connection between the problem studied here and the growth rate of black holes, in which a huge \ac{DM} can dramatically change the dynamics of the accretion flow. It is mentioned as the accretion of hot gas in galaxies' centers embedded in \ac{DM} halo.

Future surveys will uncover millions of new \acp{AGN}, leading to significant advancements in understanding these phenomena. This will allow us to test and refine models across different cosmic epochs, revolutionizing our knowledge of \acp{AGN} and their environments. The studies will also unveil the underlying physical processes that drive the growth of black holes and galaxies within the cosmic web of \ac{DM}.  
In future studies, we will test higher-resolution simulations that capture these phenomena.

Generally, the results from our simulations represent an important step toward obtaining realistic models of relativistic accretion flows that have been immersed in \ac{DM} halos.

\begin{acknowledgements}
      We thank M. Moscibrodzka and O. Porth for contributing to the initial idea and for the discussions. RR thanks S. Markoff, O. Porth and R. Wijers, at the University of Amsterdam, and M. Moscibrodzka, at the Radboud University Nijmegen and C. Herdeiro at the University of Aveiro, for their hospitality.
      This work is supported by the Center for Research and Development in Mathematics and Applications (CIDMA) through the Portuguese Foundation for Science and Technology (FCT - Fundação para a Ciência e a Tecnologia), references UIDB/04106/2020, UIDP/04106/2020 (\url{https://doi.org/10.54499/UIDB/04106/2020} and \url{https://doi.org/10.54499/UIDP/04106/2020}). HO is supported by the Individual CEEC program - 5th edition funded by the FCT, and acknowledges support from the projects PTDC/FIS-AST/3041/2020, CERN/FIS-PAR/0024/2021, and 2022.04560.PTDC. This work has further been supported by the European Horizon Europe Staff Exchange (SE) programme HORIZON-MSCA-2021-SE-01 Grant No. NewFunFiCO-10108625.
\end{acknowledgements}

%
%

\bibliographystyle{aa}
\bibliography{references}
\end{document}